\documentclass{aa}  
\usepackage{graphicx}
\usepackage{tabularx}

\usepackage{txfonts}
\usepackage{hyperref} 
\hypersetup{
    colorlinks=true,
    filecolor=magenta,      
    urlcolor=cyan,
    pdftitle={Sharelatex Example},
    citecolor=blue, 
}
\usepackage{xcolor}
\newcommand{\diff}[1]{{\color{black} #1}}

\newcommand{\FIG}[1]{{Fig.~}}
\newcommand{\EQ}[1]{{Eq.~}}
\newcommand{\SECT}[1]{{Sect.~}}

\usepackage[normalem]{ulem}

\begin{document} 

\title{Improved prior for adaptive optics point spread function estimation from science images: Application for deconvolution}


   \author{A. Lau
           \inst{1}
           \and
           R. JL. Fétick\inst{1,2}
           \and
           B. Neichel \inst{1}
           \and
           O. Beltramo-Martin \inst{3,1}
           \and
           T. Fusco \inst{1,2}
           }

   \institute{Aix Marseille Univ, CNRS, CNES, LAM, Marseille, France\\
               \email{alexis.lau@lam.fr}
          \and
              ONERA, The French Aerospace Lab BP72, 29 avenue de la Division Leclerc, 92322 Chatillon Cedex, France
            \and
              SpaceAble, 13-15 rue Taibout, 75009, Paris
             }

\date{}
\abstract
{Access to knowledge of the point spread function (PSF) of adaptive optics(AO)-assisted observations is still a major limitation when processing AO data.  This limitation is particularly important when image analysis requires the use of deconvolution methods. As the PSF is a complex and time-varying function, reference PSFs acquired on calibration stars before or after the scientific observation can be too different from the actual PSF of the observation to be used for deconvolution, and lead to artefacts in the final image.}
{We improved the existing PSF-estimation method based on the so-called marginal approach by enhancing the object prior in order to make it more robust and suitable for  observations of resolved extended objects.}
{Our process is based on a two-step blind deconvolution approach from the literature. The first step consists of PSF estimation from the science image. For this, we made use of an analytical PSF model, whose parameters are estimated based on a marginal algorithm. This PSF was then used for deconvolution. In this study, we first investigated the requirements in terms of PSF parameter knowledge to obtain an accurate and yet resilient deconvolution process using simulations. We show that current marginal algorithms do not provide the required level of accuracy, especially in the presence of small objects.
Therefore, we modified the marginal algorithm by providing a new model for object description, leading to an improved estimation of the required PSF parameters.}
{Our method fulfills the deconvolution requirement with realistic system configurations and different classes of Solar System objects in simulations. Finally, we validate our method by performing blind deconvolution with SPHERE/ZIMPOL observations of the Kleopatra asteroid.}
{} 

\titlerunning{Improved prior for adaptive optics point spread function estimation
from science images}
\maketitle

\section{Introduction}
One of the main challenges for ground-based astronomical observations at visible wavelengths is to mitigate the effects of atmospheric turbulence. Without any special treatment, the angular resolution of the biggest telescopes on Earth is naturally limited to an equivalent 10 - 20 cm telescope in the visible due to atmospheric turbulence, leading to a resolution no smaller than 0.5 – 1 arcsecond on the astronomical targets \citep{Fried1966}. Development in adaptive optics (AO) has improved the image quality by offering corrections to the perturbations of the atmosphere, partially restoring the angular resolution close to the diffraction limit of the biggest telescopes (e.g. \cite{Davies2012}). Nevertheless, these AO corrections are only partial \citep{Conan1994}, and the final point spread function (PSF) can have a complex shape.

For instance, if the AO system were to perfectly correct all the aberrations within the range of its deformable mirror, the PSF would be the combination of the airy function near the optical axis and the remains of the extended seeing-limited wings for focal positions above the correction range. In reality, the AO system is not perfect, and suffers from measurement noise, temporal or aliasing errors, among others. These errors impact the PSF shape within the correction range and strongly depend on wavelength, field position, and time. Consequently, the low spatial frequencies (global shape) of the observed scene are conserved but the high spatial frequencies (the smallest details or sharp edges) are strongly attenuated or even lost, resulting in blurring of the images \citep{Fusco2000}. Precise image processing is therefore mandatory to retrieve these fine details in the observed objects, and deconvolution is a process designed to restore the high spatial frequencies to or near to their original level \citep{Starck2002}.

Deconvolution requires knowledge of the PSF, and the quality of the deconvolved image directly depends on the accuracy of its PSF model \citep{Jolissaint2008, Davies2012}. For example, \cite{Fetick2019VESTA} and \cite{Kleopatra2021} used observations of asteroids from the European Southern Observatory (ESO) Large Program (ID 199.C- 0074, PI P.Vernazza, \cite{Vernazza2021}) to show that the classical method of getting reference PSFs by observing calibration stars before or after the science observations leads to unacceptable errors in more than 50\% of the cases. Among those cases, the deconvolution products resulted in strong artefacts at the asteroids' edges, highlighted by the presence of a bright corona. This situation is clearly the consequence of inaccurate reference stars, differing from the actual PSF. Given that the strength and spatial structure of the atmospheric turbulence are constantly evolving, a PSF calibrator acquired with the same instrument setting before or after the science exposure can be completely uncorrelated to the actual PSF.

To overcome this issue, \cite{Fetick2020BlindDeconv} introduced an approach whereby (i) the PSF is calculated using a simplified analytical model relying on a handful of parameters and (ii) these PSF parameters are estimated directly from the object thanks to a parametric marginal method. Once the PSF is estimated, it can then be used by standard deconvolution methods like Wiener filtering and Richardson–Lucy deconvolution \citep{Rick1972,Lucy1974}, where the PSF is fixed.

In this paper, we aim to develop a more robust method for PSF estimation and deconvolution for astronomical images.  In \SECT{}\ref{section:deconvolution_sensitivity}, we first investigate the extent to which  we need to know the PSF parameters for proper deconvolution under different signal-to-noise ratios (S/Ns). Then, based on these requirements, we explore the performance of the PSF-parameter estimation directly from the image under different configurations in \SECT{}\ref{section:Direct PSF Estimation from Science Observation}. In particular, we extend the work of  \cite{Fetick2020BlindDeconv} by considering different object shapes and AO-correction levels. We highlight the limitation of the previous method when dealing with elliptical objects, and we propose a generalisation of the process in \SECT{}\ref{section: PSD model of the object}. Finally, we illustrate the process with simulations and science observations from VLT-SPHERE in \SECT{}\ref{section:Applications: Simulated and On-Sky Data}.
\section{Requirement for deconvolution}
\label{section:requirment_deconovlution}
\subsection{Deconvolution method}
\label{section:deconvolution_method}
During the deconvolution process, we introduced regularisation terms to prevent a drastic increase in noise. $l^1$ and $l^2$ regularisation are common in image processing. The $l^1$ (total variation) regularisation scales with the gradient of the object, which preserves the edge or the high-frequency contents but is insensitive to low-amplitude noise or features compared to the $l^2$ (Wiener like) approach. The $l^2$ penalises the criterion by scaling quadratically with the object's gradient, and tends to over-smooth the edges or sharp features, as opposed to the $l^1$. We used MISTRAL \citep{Mugnier2004} because it uses the $l^2 - l^1$ norm. This norm is adapted to restore objects with both smooth features and sharp edges. The criterion of the deconvolution is defined as 
\begin{align}
\label{eqn:mistral}
J_{mistral} (\textbf{o}, \textbf{h}; \delta, \kappa) = || \frac{i  - \textbf{o} \circledast \textbf{h}} {\sigma_n} ||^2  + \delta^2 \sum_{i,j} |\frac{\nabla o_{i,j}}{\delta\kappa}| 
- \ln \left ( 1+ |\frac{\nabla o_{i,j}}{\delta \kappa}| \right ) , 
\end{align}
where $\delta$ is the threshold between $l^2$-like and $l^1$-like behaviour, $\kappa$ is a scaling factor of the object's gradient, $\sigma_n$ is the noise, and $\nabla$ is the isotropic gradient operator defined as, 
\begin{equation}
    \label{eqn:grad_nabla}
    \nabla o_{i,j} = \sqrt{ ( o_{i+1,j} - o_{i,j}) ^2 - ( o_{i,j+1} - o_{i,j})^2}
.\end{equation} 
The parameter $\delta$ acts as a threshold, where the criterion switches between a quadratic and the linear behaviour depending whether the gradient of the  object is larger or smaller than $\delta$. Therefore, the combination of $l^2 - l^1$ deconvolution is beneficial for high-angular-resolution objects with sharp edges because it cancels the penalties for the large gradients and ensures a good smoothing for the small gradients. We chose ($\kappa$ = 1, $\delta$ = 1.5) based on visual inspections of the deconvolution with the true PSF and applied the same value for our tolerance study. 

\subsection{Deconvolution sensitivity}
\label{section:deconvolution_sensitivity}
The sensitivity of the PSF parameters in deconvolution is often overlooked because there is no generic metric to describe the quality of the end product. Without access to the ground truth, it is difficult to derive a quantitative statement for the sensitivity of the results to the level of our knowledge of the PSF. While having a precise statement to evaluate the quality of the deconvolved object is almost impossible, here we aim to investigate the sensitivity of the process with respect to the input PSF parameters. This is achieved based on observation simulation, and the use of an analytical AO-PSF model. The advantage of the parametric PSF lies in the flexibility of its parameters, which can be adjusted to better estimate and quantify the impact of the PSF parameters on the final deconvolution accuracy.

\citet{Fetick2019VESTA} provides the first quantitative estimate of the  impact of our level of knowledge of the  PSF and the physical meaning of the parameters for deconvolution of ground-based AO observation of an asteroid surface. The PSF model in \citet{Fetick2019VESTA} was relatively simple, as they created synthetic PSFs using a Moffat profile. In particular, their PSF model does not include any physical parameters or the extended seeing halo. \diff{Nevertheless, they show that for PSFs with equivalent, large Full width half at maximum (FWHM) (or too low a Strehl ratio (SR)), the deconvolution induces the corona artefacts.} In other words, overestimating the FWHM or underestimating the SR eventually leads to deconvolution artefacts. Conversely, for small FWHMs,  or large SRs,  the resulting images appear underdeconvolved, with the extreme situation being one in which a dirac-PSF is used and there is no deconvolution at all. Therefore, inaccurate or insufficient knowledge of the PSF impacts the deconvolution accuracy with different effects on the final image depending on the under- or overestimation of the PSF performance.

\begin{figure}[t!]
    \centering
    \includegraphics[width=0.8\columnwidth]{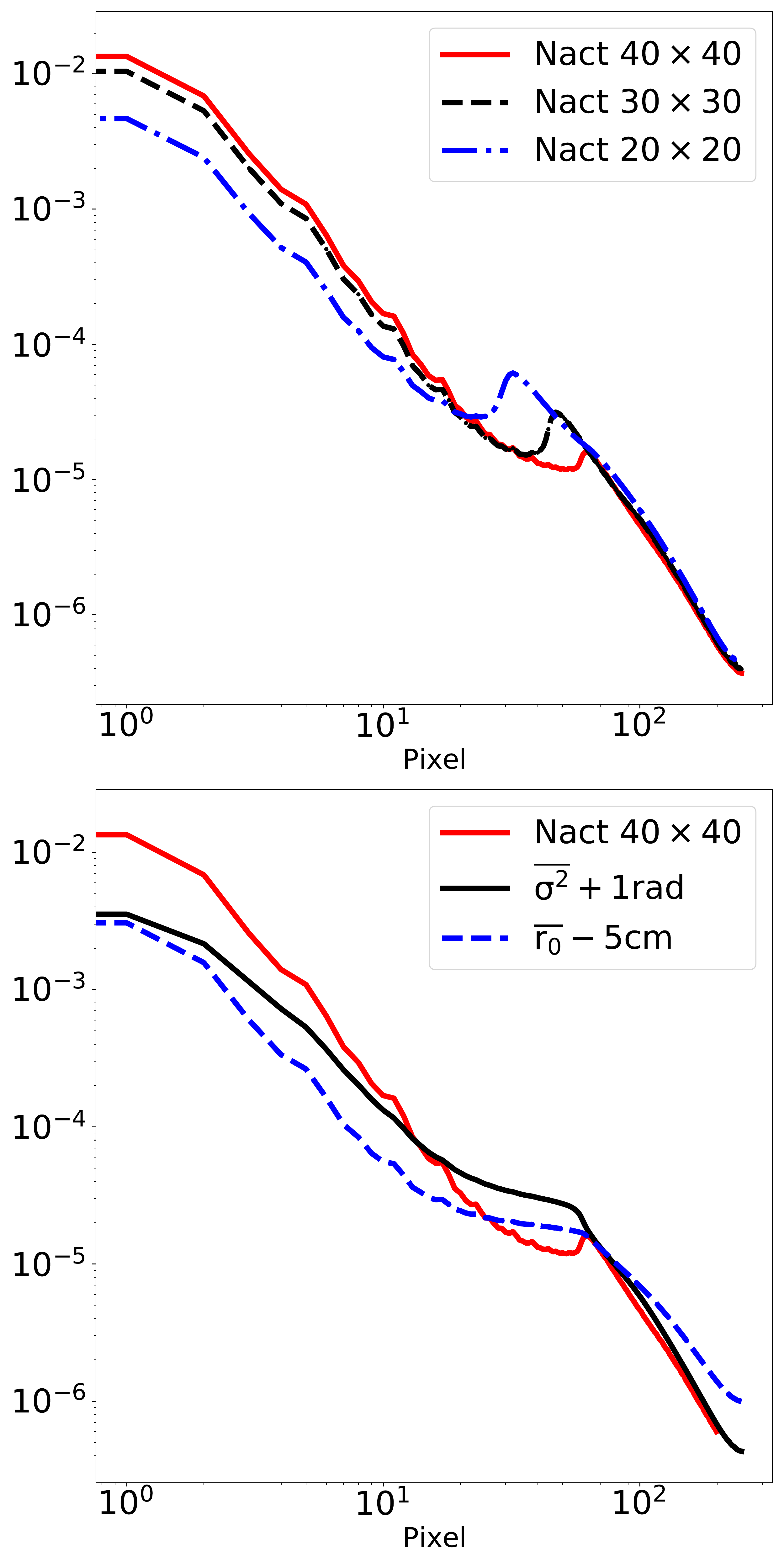}
    \caption{PSF profiles from PSFAO19 for different parameter choices. \textit{Top:} PSFs for three different values of Nact. \textit{Bottom:} PSF profiles with 40 by 40 actuators when the parameters  $r_0$ and $\sigma^2$  are changed.}
    \label{fig:psf_profile}
\end{figure}

In this work, we pursue the analysis of \citet{Fetick2019VESTA}, with a more representative and accurate AO PSF model called PSFAO19 \citep{FetickPSFAO19}. We reiterate the analytical expression of the PSF, $h$  , defined as
\begin{equation}
    W_{\phi} (f) = \left[ \sigma^2 M_A(f_x,f_y)+ C \right]_{f<f_{AO}} +  \left[{0.023r_0^{-5/3}f^{-11/3}}\right]_{f>f_{AO}}
,\end{equation}
\begin{equation}
\label{eqn:PSFAO19PSF}
h(f) = \mathcal{F}^{-1} \{{\tilde{h_T} e^{B_\phi (0)} e^{\mathcal{F}^{-1} \{ W_{\phi} (f) \}}}\},
\end{equation}
where $M_A$ is a Moffat-like function with fixed parameters, $B_\phi$ is the phase auto-correlation function, and $W_{\phi (f)}$ is the phase power spectrum density (PSD), which is characterised by physical parameters: the fried parameter ($r_0$) and phase residual variance ($\sigma^2$). Beyond the AO cut-off frequency $f_{AO}$, the PSD is simply defined by a power law (Kolmogorov spectrum), the amplitude of which is set by the fried parameter ($r_0$). The AO cut-off frequency can also be related to the number of actuators used by the AO deformable mirror (DM). Indeed, the AO DM corrects over a limited frequency range. We therefore introduce a cutoff frequency of the DM such that
\begin{equation}
\label{eqn:freqdm}
f_{AO} = \frac{1}{2 d_{DM}},
\end{equation}
where $d_{DM}$ is the inter-actuator pitch projected into the pupil, often called the DM pitch, and is defined as
\begin{equation}
\label{eqn:dmpitch}
d_{DM} = \frac{\text{D}}{\text{Nact}}
,\end{equation}
where D is the telescope primary mirror diameter and Nact is the linear number of actuators projected onto the telescope primary mirror. \FIG{}\ref{fig:psf_profile} shows different PSF profiles, as expected for an AO system installed on an 8m telescope working at visible wavelengths, when Nact changes from 20 to 40 actuators across the pupil (top figure), and when $r_0$ or $\sigma^2$ change for a given Nact.

We then took this PSF model with the parameters from Table~\ref{tab:PSF_config} to simulate science observations. We simulated observations of one of the moons of Jupiter, \object{Ganymede}, with a reference PSF generated by PSFAO19 model and a high-resolution object taken from \cite{JUNO}, and eventually binned down to fit the angular size expected from the ground, and the equivalent angular resolution for an 8m telescope working at visible wavelengths. Finally, we performed $l^2 - l^1$ deconvolution \citep{MISTRAL2003, Mugnier2004} with 225 different PSFs ---varying $r_0$ and $\sigma^2$ with respect to the true PSF parameters--- in order to carry out the deconvolution tolerance study. \\
We use the rms error (RMSE) to evaluate the performance of the deconvolution with respect to PSF parameters.  The RMSE is defined as 
\begin{equation}
\label{eqn:RMSE}
    {RMSE}_i = \sqrt{ \frac{1}{n}  \sum {(\bar{o_{i}} - \bar{o}_{ref}})^2}, 
\end{equation}
where $n$ is the number of pixels, $\bar{o_i}$ is the deconvolved object with 225 PSFs, and $\bar{o}_{ref}$ is the deconvolved object with respect to the reference PSF.
\\
\begin{table}[!h]
\centering
    \caption{System configuration for our PSF model}
    \begin{tabularx}{\columnwidth}{lll}
    \hline
    Parameter & Value & Unit \\  
    \hline\hline
    Primary mirror diameter (D) & 8 & m   \\
    Secondary mirror diameter (D') & 1.12 & m \\
    Plate scale (res) & 4.7 & mas \\ 
    Number of actuator (Nact) & $40\times40$ & \textbackslash A \\ 
    DM pitch ($d_{DM}$) & 20 & cm \\ 
    \hline
    Fried parameter ($r_0$) & 13.5 & cm\\
    Phrase variance ($\sigma^2$) & 1.3  & ${\mathrm{rad}}^2$ \\
    \hline
    \label{tab:PSF_config}
\end{tabularx}
\tablefoot{Fixed telescope parameters (top) and the reference PSF parameters (bottom).}
\end{table}

\begin{figure*}[!t]
    \centering
    \includegraphics[width=0.9\textwidth]{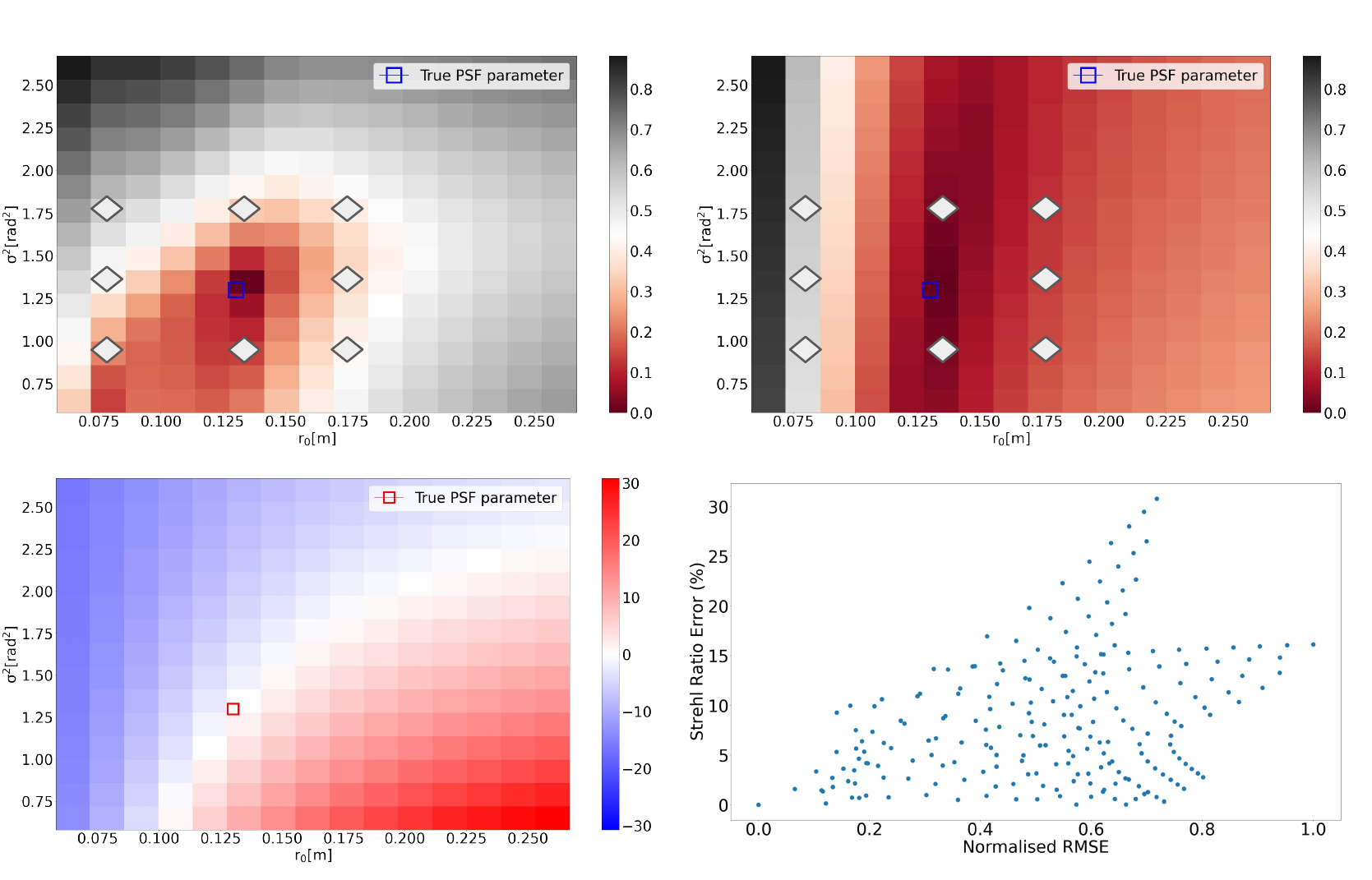}
    \caption{Deconvolution sensitivity with two different S/Ns. \textit{Top left}: Deconvolution RMSE with high S/N = 100. \textit{Top right}: Deconvolution RMSE with low S/N = 10. The colour intensity represents the strength of the RMSE. Darker colours correspond to lower RMSE values. The square represents the true PSF parameter and the diamond symbols represent the visual results displayed in \FIG{}\ref{fig:deconvolution_RMS_obj}. \textit{Bottom left:} Strehl ratio error with the PSF parameters in the deconvolution RMSE map. \textit{Bottom right}: Strehl ratio error versus normalised RMSE in the deconvolution RMSE map with a high S/N.}
    \label{fig:deconvolution_RMS}
\end{figure*}

Table \ref{tab:PSF_config} sets the reference PSF parameters for the simulations, with $r_0$ = 13.5 cm and $\sigma^2$ = 1.3 $\mathrm{rad^2}$. We choose these PSF parameters as they are representative of the typical turbulence statistics and  performance of the SPHERE extreme AO system (SAXO). 
\FIG{}\ref{fig:deconvolution_RMS} shows the RMSE maps when the PSF parameters ($r_0$,$\sigma^2$) used for deconvolution are changing. Each RMSE on the map is the sum over the pixel of \EQ{}(\ref{eqn:RMSE}). As expected, the RMSE between the real object and the deconvolved image is minimal when the exact same PSF is used for deconvolution. When the PSF parameters are changing, the deconvolution is degraded and the RMSE increases. We considered two cases of S/N, with a high S/N(=100) case (\FIG{}\ref{fig:deconvolution_RMS} \textit{left}) and a low S/N(=10) case (\FIG{}\ref{fig:deconvolution_RMS} \textit{right}), where the S/N is defined per pixel over the object. From \FIG{}\ref{fig:deconvolution_RMS}, it is interesting to note that (1) the slope of the RMSE as a function of $r_0$ is steeper than $\sigma^2$ regardless of whether there is over- or underestimation of the parameters, and (2) the RMSE is asymmetric with respect to the PSF parameters. Overestimating $r_0$ and underestimating $\sigma^2$ will always result in a better deconvolution than the opposite scenario. Our results agree with the findings of \cite{Fetick2020BlindDeconv}, as increasing $r_0$ or reducing $\sigma^2$ corresponds to a better-quality PSF. This leads to underdeconvolution, whereas overdeconvolution results in strong artefacts, which increases the RMSE dramatically. 

In order to quantify the effects described above, \FIG{} \ref{fig:deconvolution_RMS} (\textit{bottom right}) shows the normalised RMSE values for different PSFs selected over the 225 sample, for which we also computed their SR. The conclusion remains as before; if the PSF is barely known, it is always more desirable to overestimate the PSF quality before using it for deconvolution. \FIG{} \ref{fig:deconvolution_RMS} (\textit{bottom row}) also shows that using the SR as a unique metric for quantifying the PSF quality is insufficient for deconvolution, as one can have many PSFs with the same SR but very different RMSE outcome. In other words, the PSF accuracy cannot be constrained with the SR-estimation accuracy alone, as the morphology of the PSF core (in our case constrained by $\sigma^2$) also impacts the deconvolution. On the other hand, using the proposed analytical PSF model with constraints on $\sigma^2$ and $r_0$ seems appropriate for obtaining constraints on the required PSF accuracy for deconvolution. Finally, it is important to highlight that the \diff{overall tendency of the deconvolution sensitivity is independent of the S/N; therefore this latter conclusion to be applied during the PSF estimation procedure, as presented below.}

\begin{figure*}[!h]
    \centering
    \includegraphics[width=0.85\textwidth]{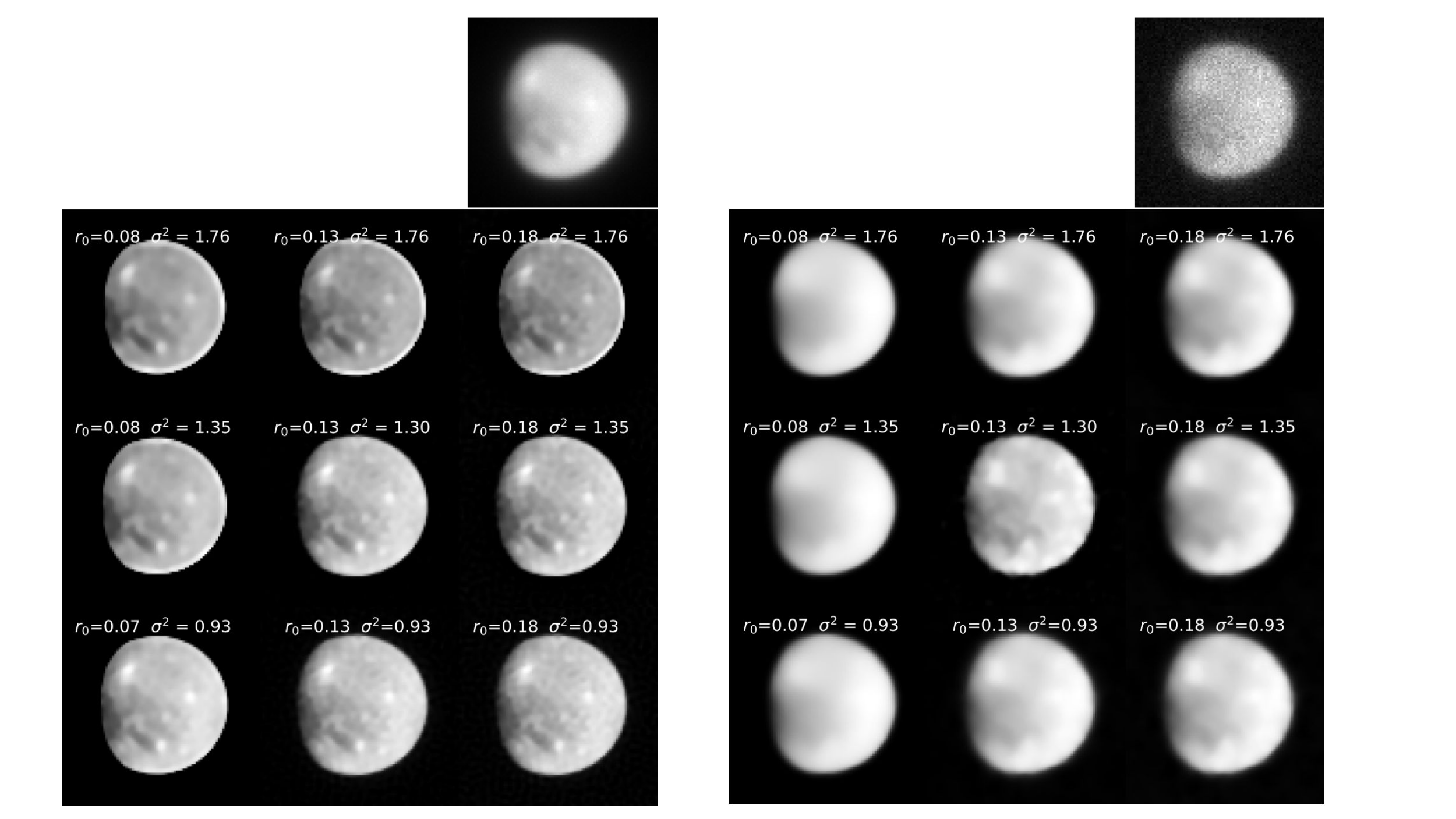}
    \caption{Simulated images produced from deconvolutions of the objects shown schematically in \FIG{}\ref{fig:deconvolution_RMS}. The simulated image before deconvolution is shown in the top right corner of each panel of nine images. \diff{\textit{Left}: Deconvolved objects marked by the diamond symbols in the \textit{top left} of fig. \ref{fig:deconvolution_RMS} with high S/N(=100), \textit{ Right}: Deconvolved objects marked by the diamond symbols in the \textit{top right} of fig.\ref{fig:deconvolution_RMS} with low S/N(=10).
    \label{fig:deconvolution_RMS_obj}}}
\end{figure*}

The final step in assessing deconvolution quality is a visual inspection, as normally there is no ground truth to compare with science observations. We inspect the impacts of over- and underestimating the PSF parameters on the deconvolution in \FIG{}\ref{fig:deconvolution_RMS_obj} and their respective RMSE are labelled by diamonds in \FIG{}\ref{fig:deconvolution_RMS}. Again with overestimated PSF parameters, we obtain better visual results in the deconvolved objects. The RMSE map (\FIG{}\ref{fig:deconvolution_RMS}) and its visualisation (\ref{fig:deconvolution_RMS_obj}) demonstrate that, in general, in order to achieve a good deconvolution, the estimated PSF parameters may fall in the ranges 
$-15\% < \mathrm{r_0} < 30\%$ and $-30\% <\mathrm{\sigma^2}<15\%$. Though these results do not define or set the generic requirement for deconvolution sensitivity, they provide quantitative guidelines as to the accuracy required on PSF parameters in order to achieve good-quality deconvolution.

\section{Direct PSF estimation from science observation}
\label{section:Direct PSF Estimation from Science Observation}
One of the traditional methods for blind deconvolution is joint-estimation \citep{Markham1999, Mugnier2004}, whereby simultaneous estimations of the PSF and the object of the image are performed. The difficulty with this method comes from the fact that we have twice as many unknowns (object and PSF) as data points, which is a very poor statistical contrast \citep{Blanco2011}, and this can end up in coupling between the PSF and the object. Another approach, called marginal estimator \citet{Blanco2011}, is to separate the object $o$ from the problem by integrating over the probability of the object distribution. This estimator is consistent and non-biased \citep{lehmann2006} when there are more data or the S/N increases. \cite{Fetick2020BlindDeconv} reduces the statistical contrast by PSF parmeterisation,
\begin{equation}
   \label{eqn:marginal_likelihood}
    P(\gamma|i) = \int P(i|o, \gamma) P(\gamma) P(o) do 
,\end{equation}
where $\gamma$ are the PSF parameters. To compute the marginalisation integral, we assume a Gaussian prior probability for the object and assume that the total noise is white (homogeneous overall), stationary, and follows Gaussian statistics.

AO images are dominated by photon noise and read-out noise. \cite{Mugnier2004}  showed that the Poisson distribution may be approximated by Gaussian noise when the flux is typically > 10 photons/pixel. The read-out noise is Gaussian, and dominates the image when the S/N is low, and so in the end, one can state that the noise is Gaussian over a wide range of S/N. The stationary distribution is a strong assumption, as when photon noise dominates (for high-S/N cases), the spatial distribution of the noise will follow the object intensity. However, these assumptions are mandatory in order to write a simple analytical expression in the Fourier domain for minimisation, and we demonstrate in section \ref{section: Limitation of the current method on PSF retrieval} that they do not introduce bias in the results. Along with the Gaussian statistics of the object, the probability function is
\begin{equation}
    \label{eqn:marg_probability}
    P(i|\gamma) \propto \frac{1}{\sqrt{\det{R_i}}} \exp^{-\frac{1}{2}(i-i_m)^t {R_i}^{-1} (i-i_m)},
\end{equation}
where $R_i$ is the image covariance matrix and $i_m$ is the convolution of estimated PSF and the mean object. With the assumption of the Gaussian stationary noise, the image covariance matrix is
\begin{equation}
    \label{eqn:image_covariance}
    R_i = H_\gamma R_o {H_\gamma}^t + \left\langle{\sigma_n}^2\right\rangle I_d,
\end{equation}
where $<\sigma_n^2>$ is the average noise variance of the pixels, $I_d$ is the identity matrix, and $H_\gamma$ is the convolution operator. The marginal criterion $J$ can be written in the Fourier domain by taking the logarithm of $P(\gamma|i)$ \citep{Blanco2011}, 

\begin{align}
    \label{eqn:marginal_criterion}
    J_{marg} (\gamma; S_{obj}, <\sigma_n^2>) = \frac{1}{2} \sum \ln{\left( S_{obj}|\tilde{h_\gamma}|^2 + <\sigma_n^2>\right)}\\ \nonumber
    +\frac{1}{2} \sum \frac{|\tilde{i} - \tilde{h}\tilde{o_m}|^2}{S_{obj} |\tilde{h}|^2 + <\sigma_n^2>} - \ln(\gamma),
\end{align}
where $S_{obj}$ is the spatial power spectrum density (PSD) of the object and $\ln{\gamma}$ represents the prior information on the PSF parameters (if there is any). This method requires prior information on the object PSD, which can be provided via an axis-symmetric model \citep{Conan1998}, 
\begin{equation}
    \label{eqn:PSD1Dmodel}
    S_{obj}= \frac{k}{1 + \left( f / \rho_0 \right)^p}, 
\end{equation}
where $k$ is the object PSD value at $f = 0$ (which is close to the square of the flux of the image), $\rho_0$ is a parameter related to the size of the object, and $p$ is the index of the power law. This object PSD model is called `PSD Conan' in the remainder of the paper. The validity of this approach was demonstrated by \cite{Fetick2020BlindDeconv}, who showed that a fully unsupervised algorithm (estimation of both the PSF and the PSD parameters of the  object) could provide reasonable-quality PSFs for deconvolution of large objects (e.g. asteroid \object{Vesta}). The authors also show that the PSF estimation can be improved with mostly unsupervised approaches, where $p$ is fixed, and all the remaining parameters are `free' during the minimisation. We estimate the parameters by minimising \EQ{}\ref{eqn:marginal_criterion} with the variable metric method (VMLM-B) descent algorithm developed by \cite{Thiebaut2002}. The PSF model and object PSD model with their respective analytical gradients are provided to the algorithm for the minimisation process. 

\subsection{Limitation of the current method on PSF retrieval}
\label{section: Limitation of the current method on PSF retrieval}
\begin{figure}[h!]
    \centering
    \includegraphics[width=0.8\columnwidth]{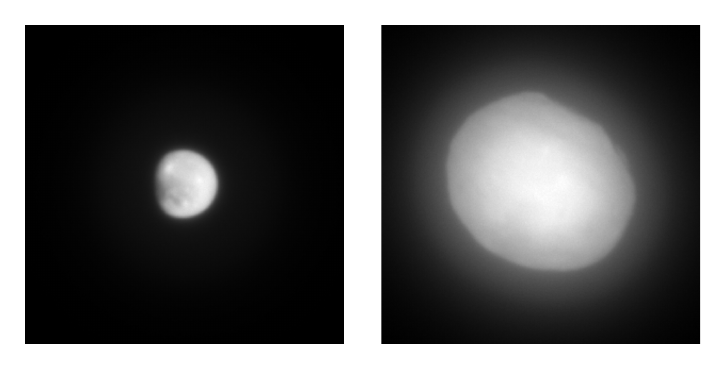}
    \caption{Simulated observations of our study. \textit{Left}: Simulated observation of \object{Ganymede}. \textit{Right:} Simulated observation of the asteroid \object{Vesta}. These observations are shown with arrays of equal size, namely 300$\times$300 pixels.}
    \label{fig:simulated_image}
\end{figure}

\begin{figure}[t!]
\centering
    \includegraphics[width = \columnwidth]{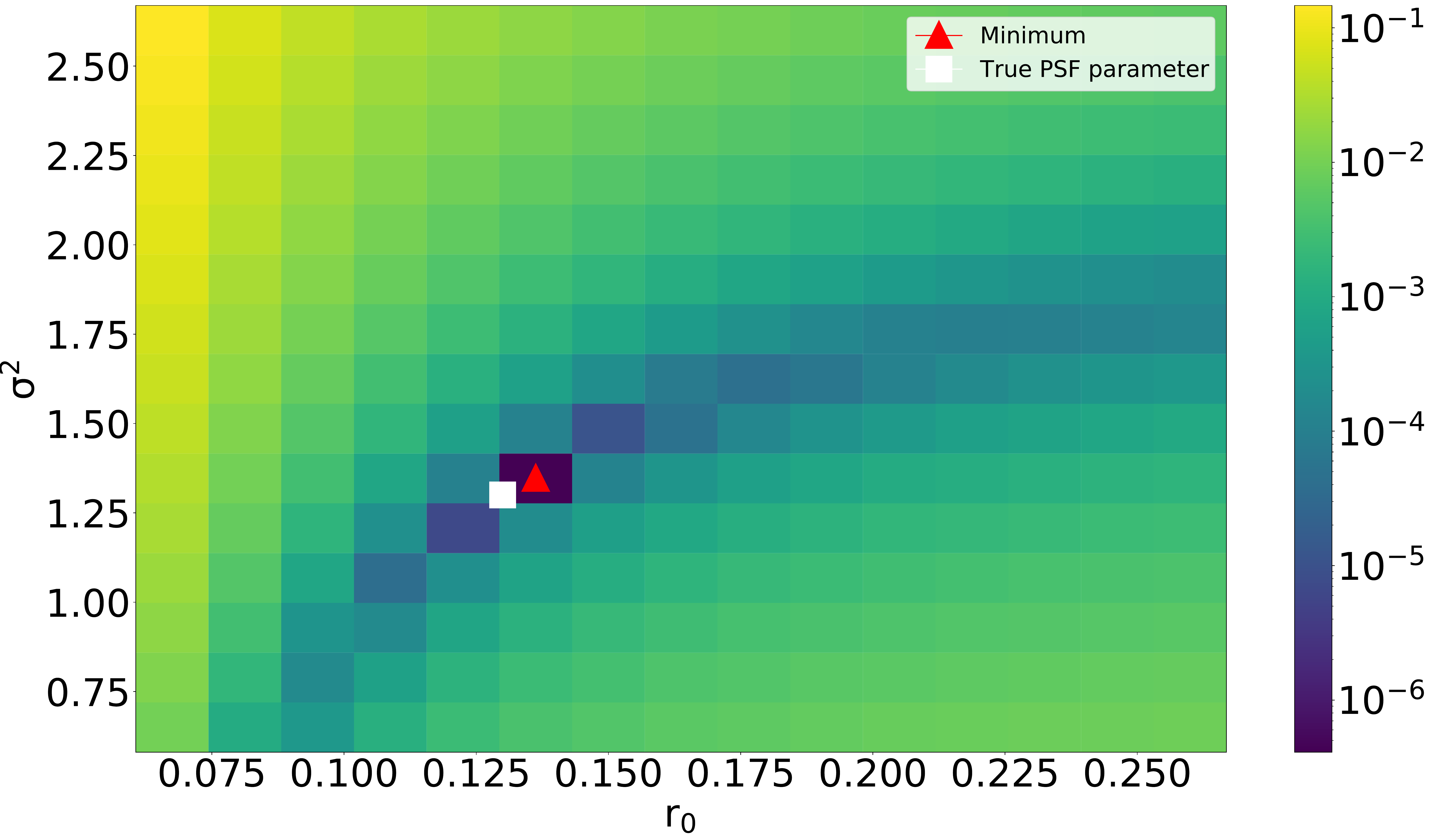}
    \caption{Supervised criterion map of \object{Ganymede} with 40$\times$40 actuators, which is the worst case from \FIG{}\ref{fig:criterion_map_1D}. We compute the map with $||F(obj)||^2$ and noise from the simulation.}
    \label{fig:Ganymede_PSD_object_map}
\end{figure}

Based on the case study of \object{Vesta,} we began including cases with less information for PSF estimation (i.e. smaller objects) and with higher-order AO corrections. \FIG{}\ref{fig:simulated_image} shows two different objects of interest, \object{Ganymede} when it is almost 100 $\%$ illuminated (case 1), and \object{Vesta} (case 2) with a larger angular size than \object{Ganymede} on the image. We use these two examples to explore the effect of different angular size on the performance of the method proposed by \cite{Fetick2020BlindDeconv}. We generated these cases with both Poisson (photon) noise and Gaussian (read-out) noise as expected in real data; this is different from the Gaussian noise assumption for the criterion \EQ{}\ref{eqn:marginal_criterion}. By doing so, we test whether the Gaussian noise assumption is valid within this marginal framework to estimate the PSF parameters. The telescope and the PSF configuration of the simulations are provided in Table \ref{tab:PSF_config}.
Before directly applying the object PSD model with the marginal estimator, we tested the robustness of the estimator by introducing the true object and noise into the model in \FIG{}\ref{fig:Ganymede_PSD_object_map}.  The supervised criterion map (or supervised map) means that all parameters are fixed, and we compute the value of the criteria for different PSF parameter values, and different pairs of $r_0$ and $\sigma^2$. \FIG{}\ref{fig:Ganymede_PSD_object_map} shows that the shape of the marginal criterion is not convex, but it seems to exhibit only one minimum. A gradient-descent algorithm is therefore suitable for solving this criterion as long as it is fine-tuned to converge precisely in the sharp valley. In the case of \FIG{}\ref{fig:Ganymede_PSD_object_map}, we substitute the object PSD model with the  modulus of the Fourier transform of the object  squared, $S_{obj}=||F(obj)||^2$. In this ideal case, the minimum of the criteria map indeed corresponds to the actual PSF parameters. In other words, if we had perfect knowledge of the object PSD characteristics, we would be able to perfectly estimate the PSF from the image. Therefore, it is crucial to provide the estimator with an accurate object description to retrieve the PSF parameters for deconvolution. This example also shows that the assumptions made in terms of the noise statistics (Gaussian) and spatial distribution (stationary noise) are not biasing the criterion, and validates these required simplifications.\\

The next step is to reproduce the same analysis in \cite{Fetick2020BlindDeconv}, computing the supervised maps for all objects in Fig. \ref{fig:simulated_image} but using the axis-symmetric object PSD model from \EQ{}\ref{eqn:PSD1Dmodel}. The object PSD parameters being fitted empirically with the data. 

\begin{figure*}[h!]
    \centering
    \includegraphics[width=\textwidth]{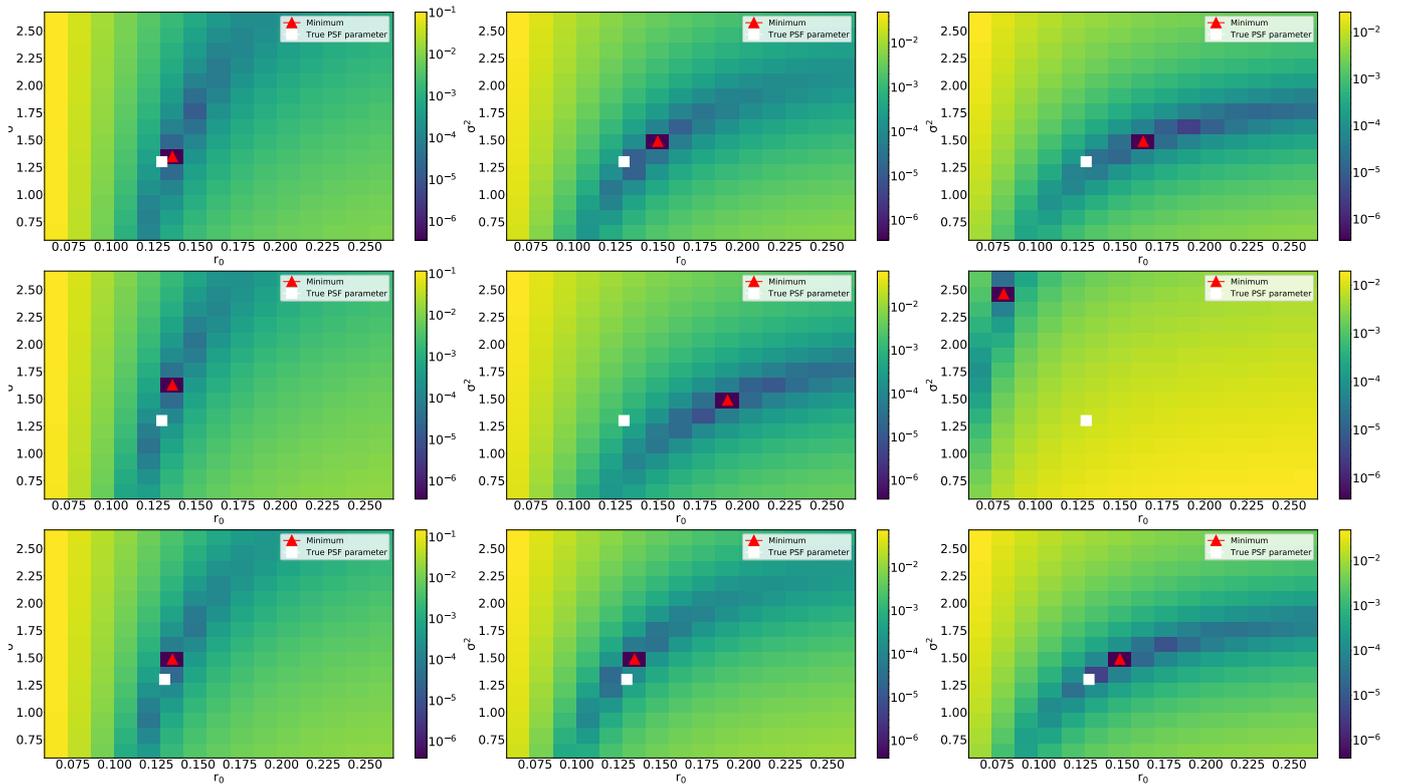}
    \caption{Supervised criterion map using the axis-symmetric and elliptical object PSD model. \textit{Top}: Vesta with the Conan model (S/N=100) [20 by 20, 30 by 30, 40 by 40]. Middle: Ganymede with the Conan model (S/N=100) [20 by 20, 30 by 30, 40 by 40]. \textit{Bottom}: Ganymede with the elliptical model (S/N=100) [20 by 20, 30 by 30, 40 by 40]. }
    \label{fig:criterion_map_1D}
\end{figure*}
The supervised maps computed for Ganymede (case 1) and Vesta (case 2) are shown in \FIG{}\ref{fig:criterion_map_1D}. Compared with \FIG{}\ref{fig:Ganymede_PSD_object_map}, the estimated PSF parameters (minimum of the supervised map) can significantly differ from the actual PSF parameters. In the case of Vesta (first row of \FIG{}\ref{fig:criterion_map_1D}), we see that when the number of actuators increases, the estimated parameters slowly diverges away from the true solution. In this case, the situation is not dramatic, as the worst $r_0$ estimated would still fall within the required sensitivity for deconvolution as defined in \SECT{}\ref{section:deconvolution_sensitivity}. 
We then look at case 1 (Ganymede), which represents a more challenging configuration, with less illuminated pixels (and therefore less information) across the object than for \object{Vesta}. Results are shown in the second row of \FIG{}\ref{fig:criterion_map_1D}. We clearly see that the PSF-parameter estimation fails, as the minimum of the criteria map notably deviates from the actual PSF parameters. At the lowest number of actuators (20 $\times$ 20), $\sigma^2$ is overestimated by 13 $\%$, which would still be acceptable for deconvolution purposes. However, when the AO-correction level is improved (Nact with a higher actuator density), \diff{the PSF-parameter estimation strongly diverges from the simulated input PSF-parameters}. For both cases, it is important to remember that if one uses the real object PSD representation (and not the axis-symmetric approximation), then the PSF-parameter estimation is accurate, as demonstrated in \FIG{}\ref{fig:Ganymede_PSD_object_map}. This is clear evidence that the source of error lies in the object PSD description, and that the axis-symmetric model assumption used in \cite{Fetick2020BlindDeconv} does not provide a proper object representation for these cases.

We interpret our finding that the order of correction impacts the PSF-parameter estimation as follows: the higher the correction, the higher the spatial frequencies corrected, and therefore the higher the energy within the AO-corrected area (less energy is `lost' in the halo). With the higher correction, the mismatch between the object PSD model and the object PSD is amplified, as more energy is translated into the object instead of the residual wavefront. A straight-line power law may no longer be sufficient to describe the features in the spatial frequencies domain. Similarly, when the angular size of the object  is too small, only a few pixels are available from the image to estimate the PSF, and the process becomes very sensitive to the assumption made on the object PSD.
 
In conclusion, even though \citet{Fetick2020BlindDeconv} further reduced the dimensionality of the problem from $2N^2 +$ to $N^2$ + the number of PSF parameters + the number of object PSD model parameters by introducing the analytical PSF model into the estimator, our study shows that with the current setup of the marginal estimator and the object PSD model (axis-symmetric), the PSF parameters cannot be estimated when (1) the object (or its angular size) is too small and (2) the order of AO correction is too high. Understanding these limitations, one may improve the object PSD model to ensure a better description  for the marginal estimator.

\subsection{Improved PSD model for the object}
\label{section: PSD model of the object}
In \SECT{}\ref{section: Limitation of the current method on PSF retrieval}, we show that the description of the object PSD prevents the marginal estimator from providing a more \diff{realistic} PSF evaluation, as the estimated minimum diverges from the simulated parameters. The estimator converges towards the true PSF parameters when the ground truth of the object and the noise are provided, as opposed to the case of Ganymede in \FIG{}\ref{fig:criterion_map_1D} with 40 $\times$ 40 actuators. A superior object PSD model should enable the estimator to perform in a more adaptive fashion, for instance, \diff{allowing higher-order AO correction and fewer pixels across the object.}
\begin{figure}[th!]
    \centering
    \includegraphics[width=\columnwidth]{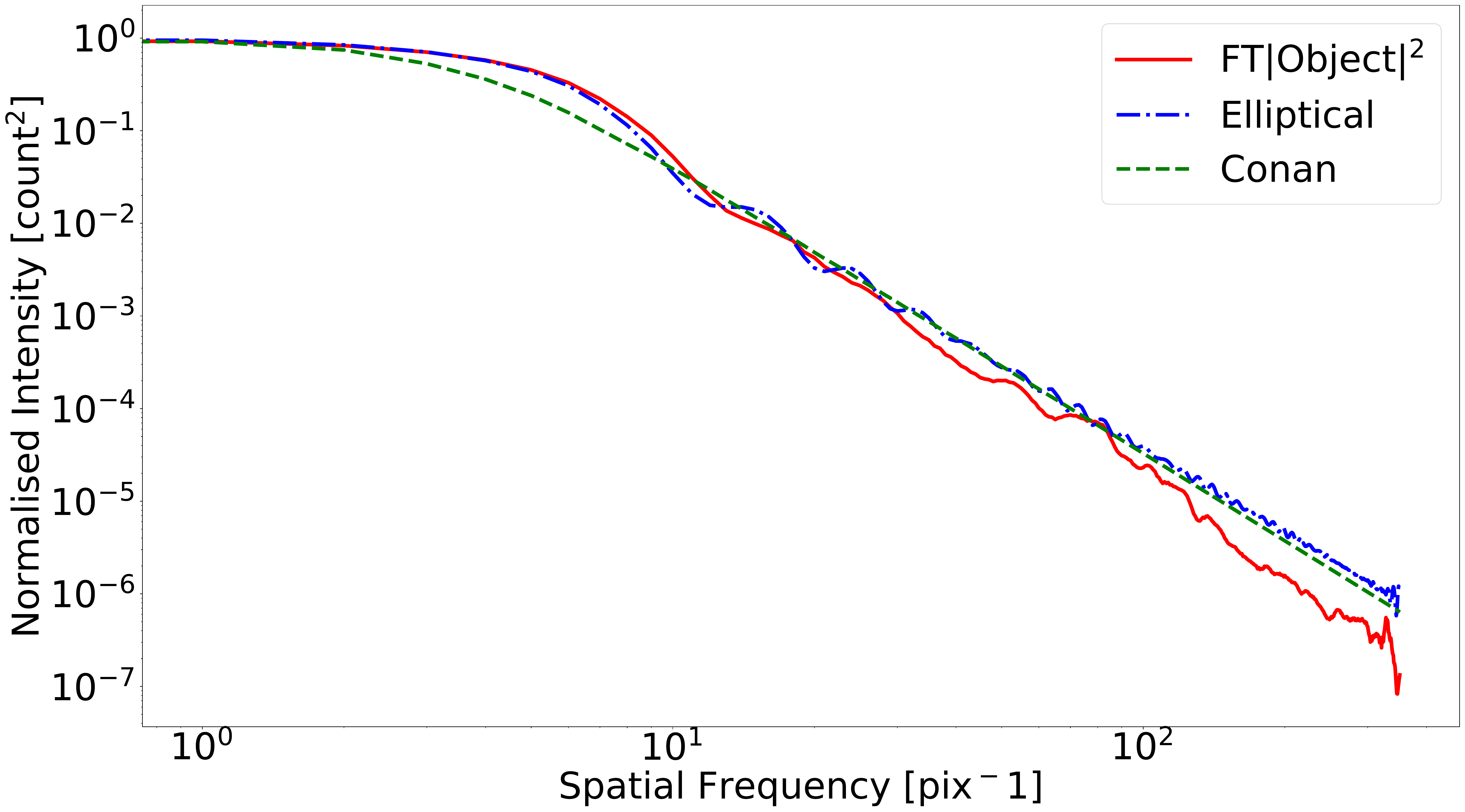}
    \caption{Circular average of the normalised intensity versus spatial frequency of $||F(obj)||^2$ (\textit{red solid line}), the elliptical object model by \EQ{}\ref{eqn:objectPSD} (\textit{blue dotted-dash line}), and the Conan model (\textit{green dashed line}).}
    \label{fig:PSD_model}
\end{figure}

\FIG{}\ref{fig:PSD_model} shows the PSD circular
average for \object{Ganymede}. From \FIG{}\ref{fig:PSD_model}, one can already see that the axis-symmetric object PSD model fails to describe the ring-like characteristics, as well as some of low-spatial-frequency behaviour, which might explain the breakdown of the marginal estimator in \FIG{}\ref{fig:criterion_map_1D}. Therefore, we propose the introduction of an elliptical function to describe the object PSD, defined as 

\begin{equation}
\label{eqn:FftEllipse}
\begin{split}
\rho^2 =  &(a^2 \sin^2 \theta + (r\times a)^2 \cos^2 \theta) f_x^2  \\&+2 ((r a)^2 - a^2) \sin \theta \cos \theta f_x f_y\\&+(a^2 \cos^2 \theta + (r\times a)^2 \sin^2 \theta) f_y^2
\end{split}
,\end{equation}
where $r$ is the radius of an ellipse, $a$ is the semi-major axis, and $r\times a$ is the semi-minor axis. We choose an elliptical model as it is more general than a simple disk, and will better represent  observations of extended sources in the Solar System. The final elliptical object PSD model can eventually be described by
\begin{equation}
    S_{obj}(\rho) = k \times |\frac{J_1 (2 \pi \rho)}{2 \pi \rho}|^2 
    \label{eqn:objectPSD}
,\end{equation}
where $k$ is the object PSD value at $f= 0$, $\rho$ is the spatial frequency, $J_1$ is the Bessel function of the first kind, and $S_{obj} (a,r ,\theta,\mu)$ is the object PSD model in the frequency domain. We adopt the same convention for $k$ as \cite{Blanco2011}, and define $\mu = k / \langle\sigma_n^2\rangle $. In practice, we find it necessary to add a Gaussian filter to the object PSD, and so the elliptical object PSD model is 
\begin{equation}
    S_{obj}(\rho) = k \times |\frac{J_1 (2 \pi \rho)}{2 \pi \rho}|^2 \circledast  \mathcal{G}, 
    \label{eqn:objectPSD}
\end{equation}
where $\mathcal{G}$ is a Gaussian function. The need for the Gaussian filter comes from the sharpness of the airy-like pattern. Sharp rings lead to some pixel values close to zero, driving the marginal criterion in an aberrant fashion. We find that a fixed Gaussian filter is adequate for all objects, and so this term remains constant. The Gaussian filter $\mathcal{G}$ we use is 
\begin{equation}
    \label{eqn:gaussian}
    \mathcal{G} = \frac{1}{2\pi\sigma^2} e^{-(x^2+y^2)/(2\sigma^2)}, 
\end{equation}
where $x$ and $y$ are the coordinate of the array, and $\sigma$ is the standard deviation of the distribution, the value of which is fixed to 2 for the filter.

With this particular formalism, and unlike the model in \cite{Conan1998}, the physical information for the object is compressed into the variables, with $a$ (semi-major axis) and $r$ (ratio between semi-major and semi-minor axes) being directly related to the size of the object. Here, $\theta$ relates to the orientation of the object PSD and $\mu$ is associated with the noise level of the image. Considering that this model is analytical, minimising the model parameters with their analytical gradients is beneficial to our multi-variable PSF estimation. The object PSD parameters are a secondary product of the marginal criterion because the goal is to retrieve the PSF parameters to build a PSF for deconvolution. 

\FIG{}\ref{fig:PSD_model} shows how this elliptical model improves the representation of the PSD features. The object PSD parameters are derived by fitting $||F(obj)||^2$ with the elliptical model from \EQ{}\ref{eqn:objectPSD}. The next step is to validate the performance of this PSD model for PSF-parameter estimation with simulations.

\section{Applications: Simulation and on-sky data}
\label{section:Applications: Simulated and On-Sky Data}

\subsection{Simulation: Fully supervised case}
We first test the marginal estimator with the elliptical PSD model by reproducing the analysis of \FIG{}\ref{fig:criterion_map_1D} for the most problematic case of Ganymede (small object) with the highest AO correction (bottom-right plot of \FIG{}\ref{fig:criterion_map_1D}). Compared with the bottom-right plot of \FIG{}\ref{fig:criterion_map_1D}, introducing the elliptical PSD model into the marginal estimator leads to a significant improvement, as shown in \FIG{}\ref{fig:criterion_map_1D}. The minimum of the criteria is now within less than 10\% of the actual PSF parameters. Again, this result confirms that the assumptions made in the noise model do not introduce bias into the estimation process.

For the case of \object{Ganymede}, the relevant parameters are found to be as follows:

\begin{equation}
\begin{cases}
\;r_0 &= 14.2 \:\mathrm{cm}\\ 
\;\sigma^2 &= 1.46 \:\mathrm{rad^2}\\ 
\;\mu &= 9.037\times 10^{-9}\\ 
\;a &= 34.0 \:\mathrm{pix}\\
\;r &= 0.867\\ 
\theta &= -1.48 \:\mathrm{rad.} \\
\end{cases}
\end{equation}

\diff{When the object parameters $(\mu, a, r, \theta)$ are well calibrated}, our elliptical model provides a stronger constraint towards the estimation of ($r_0$, $\sigma^2$), and therefore allows a degenerate regime to be avoided, as opposed to the axis-symmetric object PSD model shown in the bottom-right plot of \FIG{}\ref{fig:criterion_map_1D}.

\subsection{Simulation: Fully unsupervised case}
Without knowing the identity of a target object, one can perform the marginal estimator along with object parameters. This is what we refer to as `fully unsupervised', when the optimisation process is carried out on all free parameters, estimating both the object PSD parameters and the PSF at the same time.  \cite{Fetick2020BlindDeconv}  showed that such an approach does not provide satisfactory results, mostly because the criterion is flat and because of the couplings between parameters. We reproduced the experiment performed by these latter authors using the elliptical PSD model, but the same limitations are faced, especially with a strong coupling between $r_0$ and $\mu$.

In the Fourier domain, the PSD of the image is $ S_{obj} \times |\tilde{h}|^2 + n$, and both $r_0$ and $\mu$ determine the intensity of the image PSD. The estimator therefore has difficulty in separating the contribution of the PSF from the object. This means a stronger constraint is required either on the PSF parameter ($r_0$) or on the object parameter ($\mu$). In particular, we note that, if we can provide initial guesses that are sufficiently close to the true value (-5\% to +5\%), then the minimisation can converge towards a good solution. The fully unsupervised case therefore does not appear to be a robust PSF-estimation method.

\subsection{Simulation: Mostly unsupervised case}
\label{sec:mostlyunsupervised}

\begin{figure}[h!]
    \centering
    \includegraphics[width=\columnwidth]{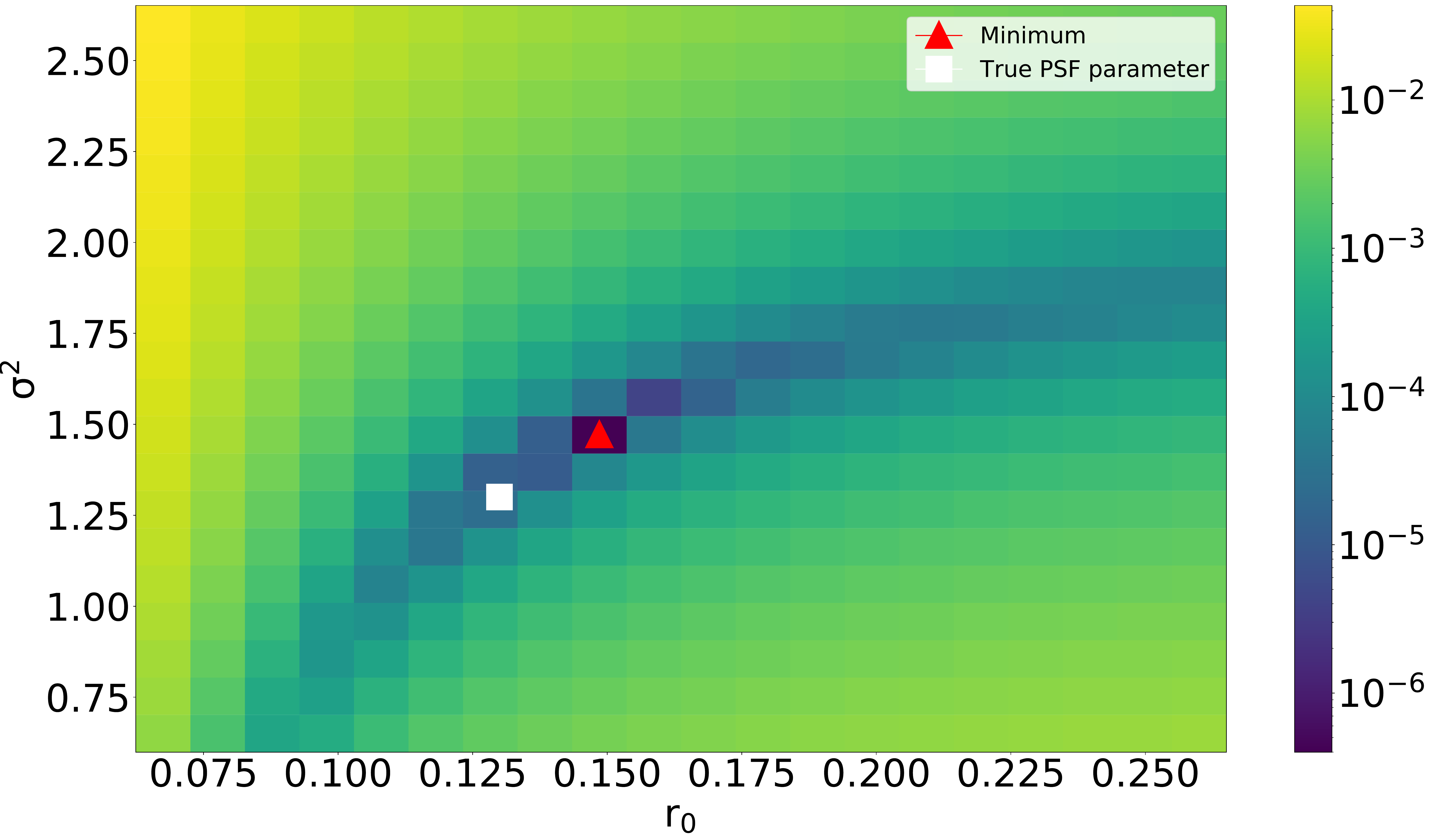}
    \caption{Mostly unsupervised criterion map of Ganymede 40 $\times$ 40 using the elliptical PSD model in \EQ{}\ref{eqn:objectPSD}.}
    \label{fig:criterion_map_mostly_unsup}
\end{figure}
For actual observations, we do have access to some information for the object from the observation itself. For instance, our object PSD model allows us to obtain relevant guesses for $S_{obj}$ via $(a,r ,\theta,\mu)$ directly from the Fourier domain of the image, which was not possible from the axis-symmetric object PSD model. Indeed, these parameters are linked to the physical parameters, which directly come from the size and orientation of the object in the image. Furthermore, $\mu$ can be estimated directly from the circular average of the image PSD by calculating the ratio between the maximum and the minimum of this circular average. Therefore, the PSF parameter estimation can be carried out with $\mu$ fixed to the guess obtained from the science image, and allowing $(a,r ,\theta)$ free to vary, starting from educated guesses. Applying such a method, \FIG{}\ref{fig:criterion_map_mostly_unsup} shows the resulting criteria maps for the same case as shown in \FIG{}\ref{fig:criterion_map_1D}, with the reference $r_0$ = 13.0 cm and $\sigma^2$ = 1.3 $\mathrm{{rad}^2}$. The relevant parameters are found to be, 

\begin{equation}
\begin{cases}
\;r_0 &= 14.9\:\mathrm{cm}\\
\;\sigma^2 &= 1.47\:\mathrm{rad^2}\\
\;\mu &= 9.03\times 10^{-9}\\
\;a &= 34 \:\mathrm{pix}\\
\;r &= 0.866\\
\theta &= -1.48 \:\mathrm{rad.}\\
\end{cases}
\end{equation}
In conclusion here, fixing $\mu$ significantly improves the PSF estimation because it removes its coupling with $r_0$. With the initial guesses of the object parameters obtained from the observation, we are able to retrieve a solution close to the fully supervised case (known object information). This approach provides a robust way of estimating the PSF parameters from the images, even in specific configurations of small and elongated objects. The final step is to validate the approach with real data, as shown in the following section.

\subsection{On-sky: Fully or mostly unsupervised}
\diff{Here, we make use of actual observations to test and validate the newly proposed algorithm. To test whether or not our marginal estimator with the elliptical object PSD model provides robust estimates of the PSF parameters, }we select a test case, namely the observation of the asteroid \object{Kleopatra} from 14 July 2017, because its elliptical shape provides an appropriate challenge to our estimator. The asteroid \object{Kleopatra} was observed as part of the European Southern Observatory (ESO) Large Program (ID 199.C- 0074, PI P. Vernazza - \cite{Vernazza2021}), targeting Main Belt asteroids to study their albedo, shape, excavation volume, and surface. This ESO Larger Program used the Zurich IMaging POLarimeter (ZIMPOL) instrument with the $N_R$ filter of the central wavelength $\lambda$ = 645.9 nm. ZIMPOL is mounted on Spectro Polarimetric High-contrast Ex-oplanet REsearcher (SPHERE) of the Very Large Telescope (VLT) observatory. It is equipped with SAXO, a high-order AO system. The images were previously deconvolved by \cite{Kleopatra2021}. In this case, we have no prior information on the object, and we apply the strategy defined in \SECT{}\ref{sec:mostlyunsupervised}: fix $\mu$ and extract first guesses for $(a,r ,\theta)$ directly from the image. We then perform PSF estimation and use this to deconvolve the image. 

For this observation, the PSF parameters found are 
\begin{equation}
\begin{cases}
\;r_0 &= 12.1 \:\mathrm{cm}\\
\;\sigma^2 &= 4.17 \:\mathrm{rad^2}\\
\;\mu &= 2.42\times 10^{-9}\\
\;a &= 29 \:\mathrm{pix}\\
\;r &= 0.55\\
\theta &= -0.92 \:\mathrm{rad.}\\
\end{cases}
\end{equation}

\begin{figure}[!ht]
    \centering
    \includegraphics[width=\columnwidth]{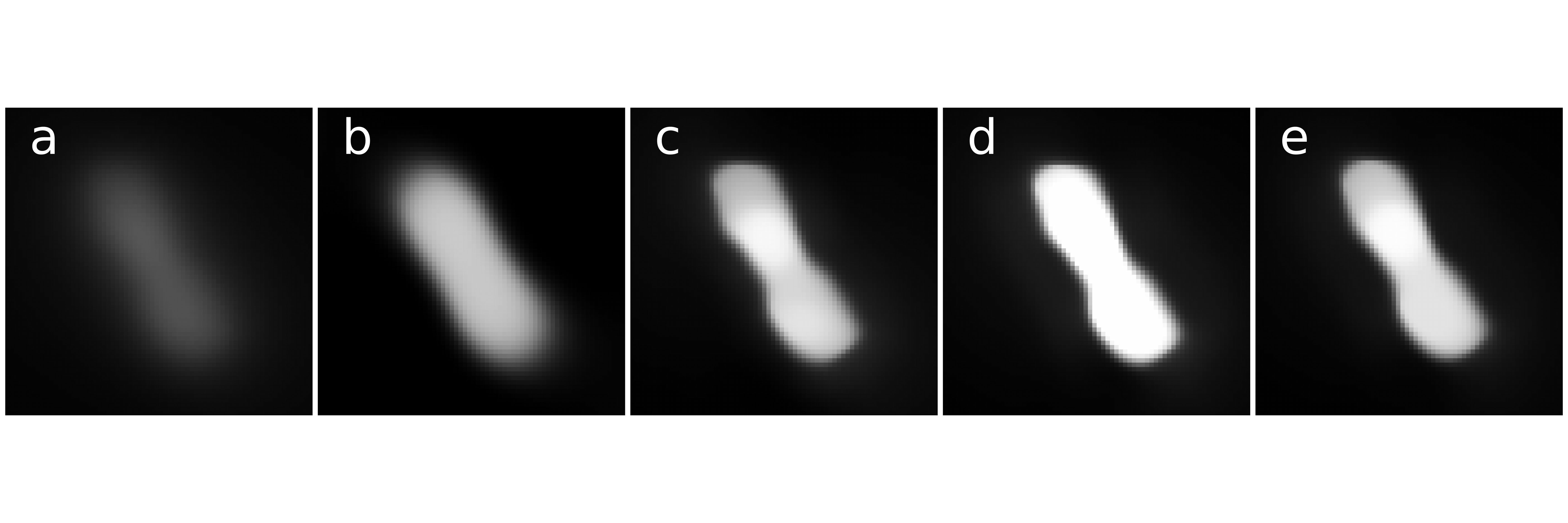}
    \caption{Observation and the deconvolutions with different PSFs. (a) Observation. (b) Deconvolution performed using the PSF estimated using a previous model \cite{Conan1998}. (c) Deconvolution using the estimation with the elliptical model. (d) Deconvolution using the calibration star observed after the observation. (e) Deconvolution by \cite{Kleopatra2021} using a Moffat PSF profile. These images are plotted with the same intensity scale.}
    \label{fig:deconv_kleo}
\end{figure}

\begin{figure}[!ht]
    \includegraphics[width=\columnwidth]{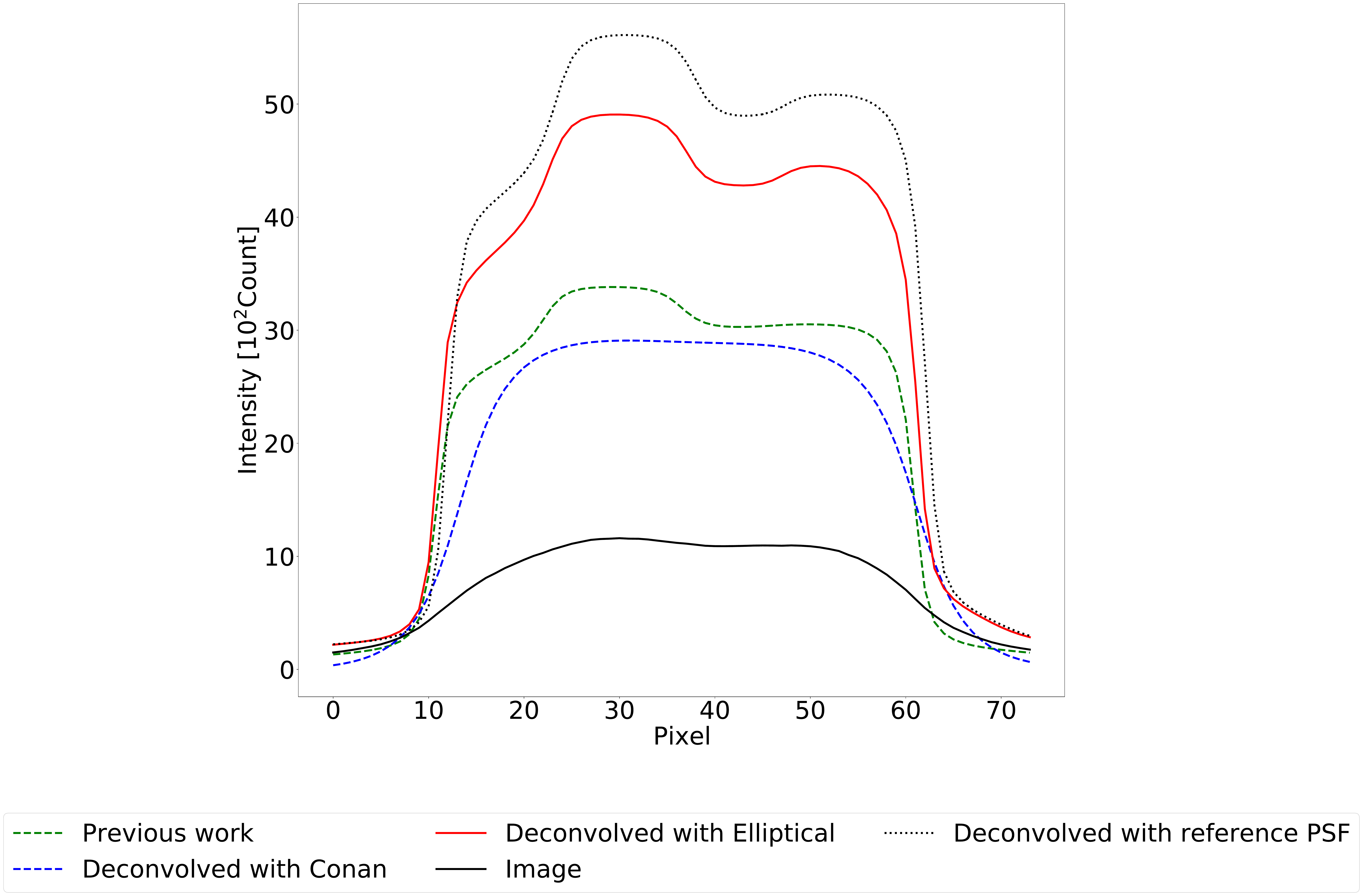}
    \caption{Intensity of the image and the deconvolutions. Slice of the asteroid \object{Kleopatra} for the image (black dotted line), previous work related to \cite{Kleopatra2021} (green star-dotted line), a deconvolution using the marginal estimator with the axis-symmetric model \cite{Conan1998} (red solid line), and a deconvolution using the marginal estimator with elliptical model (blue dash-dotted line).}

    \label{fig:deconv_kleo_intensity}
\end{figure}

\begin{figure}[!ht]
    \includegraphics[width=0.9\columnwidth]{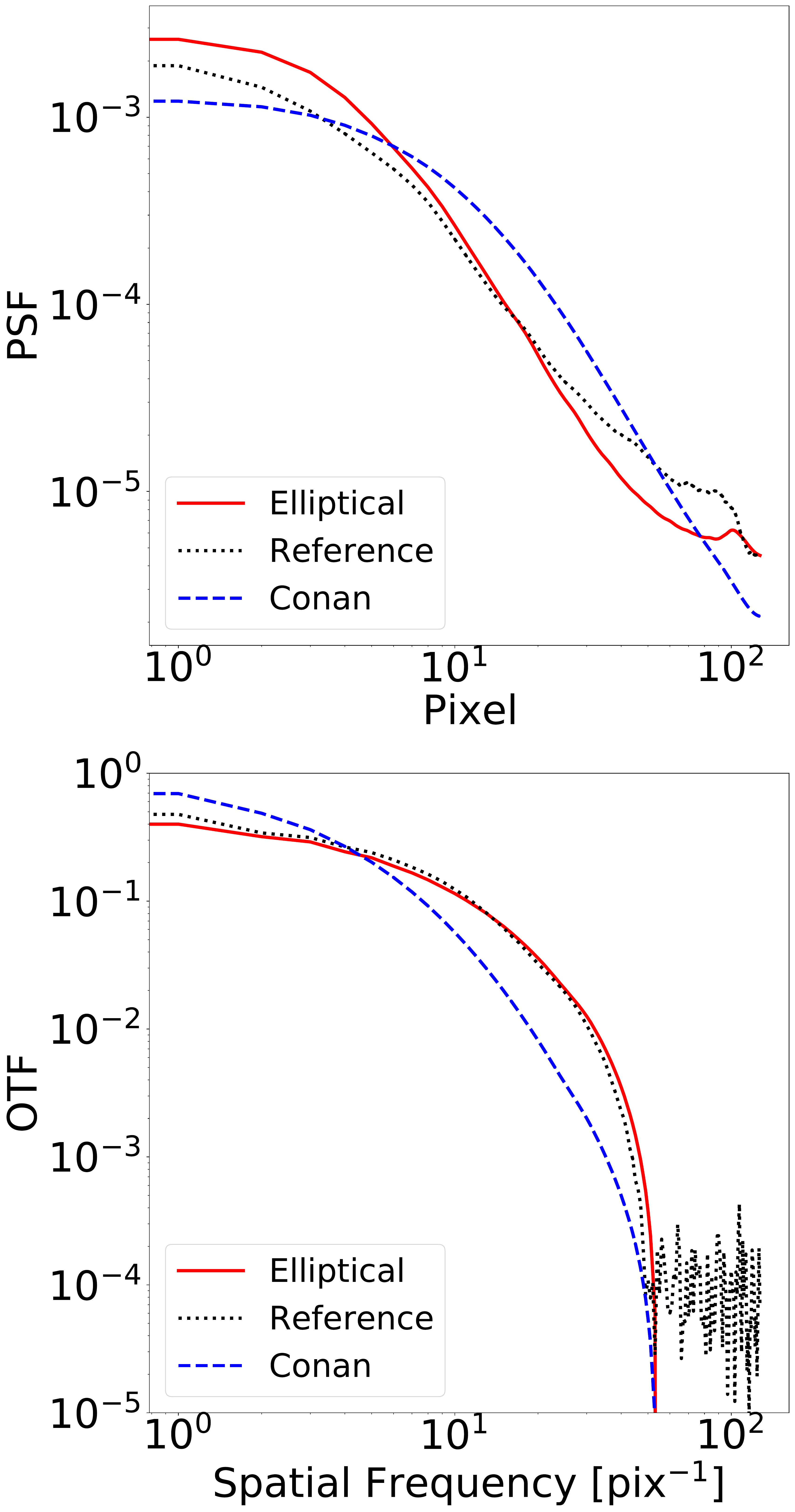}
    \caption{ PSFs and OTF profiles used for deconvolution. \textit{Top}: Cut of the PSFs and the reference PSF. The PSF estimated with by \cite{Conan1998} is shown with a red solid line, the PSF estimated with the elliptical model is shown with a blue dot-dashed line, and the reference PSF obtained after the observation  is shown as a green triangle-dotted line. \textit{Bottom}: Cut of the OTFs and the reference OTF. The OTF estimated by \cite{Conan1998} is shown as a red solid line, the OTF estimated with the elliptical model is shown as a blue dot-dashed line, and the reference OTF obtained after the observation is shown as a green triangle-dotted line.}
    \label{fig:deconv_kleo_PSF}
\end{figure}

We compare the result of our algorithm to other deconvolution strategies in \FIG{}\ref{fig:deconv_kleo}, that is, with: (a) the observation, (b) a deconvolution performed using the PSF estimated with the previous model \cite{Conan1998}, (c) a deconvolution using the estimation with the elliptical model, (d) a deconvolution using the calibration star observed after the observation, and (e) the deconvolution by \cite{Kleopatra2021}, who used a Moffat PSF profile. Regarding the deconvolution, we use a $l^1-l^2$ deconvolution scheme \cite{MISTRAL2003} while keeping all hyperparameters consistent for all the different deconvolved objects. \FIG{}\ref{fig:deconv_kleo_intensity} \diff{shows the intensity along the longest axis of the image, the deconvolution of the axis-symmetric model, the deconvolution of the elliptical model, and the deconvolution of the previous work.} The estimation made by the elliptical model successfully retrieves the morphology of the asteroid while avoiding overdeconvolution. Finally, \FIG{}\ref{fig:deconv_kleo_PSF} shows the PSF and the OTF profile of the estimated PSF with the elliptical model (blue dot-dashed line), a reference PSF (green triangle dash line) observed after the asteroid observation, and the estimated PSF with the axis-symmetric model. We can retrieve the overall PSF shape with the elliptical model when comparing with the reference PSF, assuming it provides a good proxy for the PSF.

\section{Conclusion and perspectives}
\label{section:conclusion}
This paper addresses the deployment of advanced data-processing techniques in the framework of AO-assisted observations of extended targets and Solar System objects in particular. AO correction is not perfect, especially at visible wavelengths, and image post-processing is mandatory to retrieve the finer details in the observed objects. Deconvolution is the image post-processing step to reduce the residual blurring effect on the image, but it requires an accurate description of the actual AO-PSF. In the first part of this paper, we make use of an analytical AO-PSF model to investigate the 
sensitivity of the deconvolution process to the  accuracy of the available knowledge of the PSF. We defined boundaries on the PSF-parameter values in order to achieve a deconvolution without inducing strong artefacts. Our results show that if the estimated PSF parameters fall in the ranges of --15\% < $r_0$ < 30\% and --30\% < $\sigma^2$ < 15\%, \diff{then we can achieve deconvolutions without introducing many artefacts and distortions.} The boundaries are not symmetrically distributed, and we find that it is always better to overestimate the PSF quality in order to reduce the error in the deconvolution, which agrees with the findings of \cite{Fetick2020BlindDeconv}. We also conclude that the SR alone is not a good metric for the PSF description for deconvolution.\\
We then explored algorithms allowing to extract the PSF directly from the science images. Such methods, known as marginal blind-deconvolution, have previously been shown to be efficient, especially for Solar System-like observations \citep{Fetick2020BlindDeconv}. In \SECT{}\ref{section: Limitation of the current method on PSF retrieval}, we illustrate the limitations of such a method for specific cases, and in particular for  objects of small angular size. We then improved the object description by introducing a new elliptical object PSD model in \SECT{}\ref{section: PSD model of the object} and we present the results of our tests of this elliptical model  in \SECT{}\ref{section:Applications:
Simulated and On-Sky Data}, which use the problematic cases from \SECT{}\ref{section: Limitation of the current method on PSF retrieval}. After testing with simulations, we validated our model with on-sky data from VLT-ZIMPOL. With the elliptical object PSD model, we successfully retrieve the PSF from the image and improve the shape and flux retrieval on the observations. 

Other than improving the PSF estimation with the challenging cases for the previous model, more importantly, our model allows us to obtain relevant guesses directly from the Fourier domain of the image. The object parameters required are linked to physical parameters, that can be estimated from the size and orientation of the object in the image. As such, the process is more robust as it starts from educated guesses.

In future work, a first path to improve the PSF estimation would be to use external data available from AO telemetry in order to introduce priors on the PSF parameters. AO telemetry data provide insights into observing conditions and AO performance, from which we can obtain estimations of $r_0$ and/or $\sigma^2$. These priors may be used to improve the minimisation process even further, and eventually the robustness of the PSF estimation. Another path would be to extend the method to 3D data cubes, for instance from Integral Field Spectrograph observations. Assuming that the structure of an object does not change significantly with wavelength, a single set of object PSDs could be used for all wavelengths, which would  further improve the estimation algorithm. Furthermore, one could also take advantage of the known PSF scaling with wavelengths, reducing the number of unknowns with respect to the data  even further, which would constrain the output PSF. Finally, further work is needed to adapt our method to different classes of extended objects; for example, galaxies with different topologies. Extending the blind deconvolution method to handle a more diverse class of objects will either need additional inputs from the astronomical model, or machine learning methods could perhaps be used to build more complex models. Access to the Python package of this work is possible upon request. 

%
%
\begin{acknowledgements}
This work benefited from the support of the the French National Research Agency (ANR) with WOLF (ANR-18-CE31-0018), APPLY (ANR-19-CE31-0011) and LabEx FOCUS (ANR-11-LABX-0013); the Programme Investissement Avenir F-CELT (ANR-21-ESRE-0008), the Action Spécifique Haute Résolution Angulaire (ASHRA) of CNRS/INSU co-funded by CNES, the ECOS-CONYCIT France-Chile cooperation (C20E02), the ORP H2020 Framework Programme of the European Commission’s (Grant number 101004719) and STIC AmSud (21-STIC-09).
\end{acknowledgements}

\bibliographystyle{aa} 
\bibliography{ref} 
\end{document}